\documentclass[aps,prb,tightenlines,showpacs,nofootinbib]{revtex4}
\usepackage{graphicx}
\usepackage{amsmath}
\usepackage{amsfonts}
\usepackage{amssymb}
\usepackage{bm}
\begin{document}

%\input amssymb.dtx
%\draft
\title{Simultaneous Charge Ordering and Spin Dimerization
in Quasi-Two-Dimensional Quarter-Filled Ladders}
\author{Gennady Y. Chitov}
\affiliation{Department 7.1-Theoretical Physics,
University of Saarland, Saarbr\"ucken 66041, Germany}
\author{Claudius Gros}
\affiliation{Department 7.1-Theoretical Physics,
University of Saarland, Saarbr\"ucken 66041, Germany}
\date{\today}
\begin{abstract}
We study the spin-pseudospin Hamiltonian of
the Ising Model in Transverse Field (IMTF) for pseudospins, coupled
to the $XY$-spins on a triangular lattice. This model appears from
analyses of the quarter-filled ladder compound $\rm NaV_2O_5$, and
pseudospins represent its charge degrees of freedom.
In the molecular-field approximation we find that the model
possesses two phases: charge-disordered
without spin gap; and a low-temperature phase containing both the
anti-ferroelectric (zigzag) charge order and spin dimerization
(spin gap). The phase transition is of the second kind, and the
calculated physical quantities are
as those one expects from the Landau theory.
One of particular features of the phase diagram is that the
inter-ladder spin-pseudospin coupling, responsible
for the spin gap generation, also destroys the IMTF
quantum critical point, resulting in the exponential behavior of
$T_c$ in the region of Ising's coupling where the IMTF is always
disordered.
We conclude that our mean-field results give a qualitatively correct
description of the phase transition in $\rm NaV_2O_5$, while a
more sophisticated analysis is warranted in order to take into account
the thermal fluctuations  and, probably, the
proximity of the IMTF quantum critical point.
\end{abstract}
%\pacs{05.30.Fk, 71.10Pm, 11.10.Hi, 05.10.Cc}

\maketitle
%
%
%%%%%%%%%%%%%%%%%%%%%%%%%%%%%%%%%%%%%%%%%%%%%%%%%%%%%%%%%%%%%%%%%%%%%%%%%%%%%%
\section{Introduction}\label{Intro}
%%%%%%%%%%%%%%%%%%%%%%%%%%%%%%%%%%%%%%%%%%%%%%%%%%%%%%%%%%%%%%%%%%%%%%%%%%%%%%
%
%
Properties of the models with coupled spin and orbital degrees of freedom
have been actively studied from early 70-th, mostly in the context of the
Jahn-Teller transition metal compounds.\cite{Kugel82}
In recent years it has been a growing theoretical
effort\cite{Arovas95,Ners97,Pati98,Kolezh98,Azaria,Orign00,Itoi00,
Martins00,Zhang03}
in order to understand the ground-state properties and excitation spectra
of a particular (high-symmetric) class of one-dimensional spin-orbital
models. The latter can also be viewed (up to some slight modifications at
most) as two spin-$\frac12$ Heisenberg chains coupled via a bi-quadratic
exchange. This interest is motivated by various unusual states, occurring
in many kinds of low-dimensional magnetic systems:
e.g., TDAE-$\rm C_{60}$;\cite{Arovas95} two-leg
spin-$\frac12$ ladder compounds (for a review, see Ref.[\onlinecite{Dag96}]);
recently discovered spin-gapped materials
$\rm Na_2Ti_2Pn_2O$ ($\rm Pn=As,Sb)$\cite{Pn2O}
and $\rm NaV_2O_5$,\cite{Isobe96} which do not fit into the standard
spin-Peierls scenario.

For the one-dimensional (1D) spin-orbital model there is a
special point (i.e., coupling's values) where the effective Hamiltonian
has an extra $SU(4)$ symmetry, and the exact solution has been known
for quite a long time.\cite{SU4exact} Another exactly solvable special point
for the $SU(2) \otimes SU(2)$-symmetric Hamiltonian was found recently by
Kolezhuk and Mikeska.\cite{Kolezh98} Apart of these
special values of couplings, the available results on the 1D-spin-orbital
models are known from mean-field analyses,
renormalization group, bosonization, conformal field theory, numerics,
and/or their combinations.

As was re-emphasized recently,\cite{Pati00}
from the view point of application to real materials we are mostly
interested in, high symmetry
of a spin-orbital Hamiltonian is rather artificial.
Even if the total spin is conserved from physical grounds,
i.e., the symmetry of the Hamiltonian's spin sector is indeed $SU(2)$,
the symmetry of the orbital sector is usually lower (e.g., Ising or
$XY$-type). So, applicability of the results obtained from
high-symmetric Hamiltonians (e.g., $SU(2) \otimes SU(2)$) to the cases
of low-symmetric Hamiltonians following from microscopic considerations
of a given problem, is not so straightforward.\cite{ZamNote}

It has been argued that the spin-gapped compounds
$\rm Na_2Ti_2Pn_2O$\cite{Pati98} and $\rm NaV_2O_5$\cite{Smo98}
can be modeled by a two-band quarter-filled Hubbard Hamiltonian.
In the regime of strong on-site Coulomb repulsion such
Hamiltonian can be mapped onto a spin-orbital model,\cite{Kugel82}
so the appearance of a spin gap is due to
interplay between the charge and orbital degrees of freedom.
Instead, one can work directly with the electronic Hamiltonian.
Phase diagram of the 1D quarter-filled two-band Hubbard model
(Hubbard ladder) has been recently studied both
numerically\cite{Vojta01} and analytically.\cite{Orign03}

In this paper we will explore a particular scenario for the
spin gap mechanism in a spin-orbit model, where the orbital
(``pseudospin'') degree of freedom corresponds physically to the
ordering charge displacement (disproportionation). Such a
model appears in the applications to $\rm NaV_2O_5$.\cite{PnNote}
The quarter-filled ladder-system\cite{Smo98} $\rm NaV_2O_5$ is up-to-date
the only well-established  transition-metal compound with a localized
spin-$\frac12$ moment distributed, in the high-temperature phase, equally
over two transition-metal
ions in a mixed valence state. This unique situation of
two $V^{4.5}$ ions sharing one spin $s=\frac12$ leaves room
for the charge disproportionation
$2V^{4.5}\rightarrow V^4+V^5$ occurring below the critical temperature of
$T_c=34\,{\textrm K}$, together with the opening of a spin-gap
(without a long-range order for spins). At $T_c$ the unit cell doubles both
along the ladders ($b$-direction) and along the rungs ($a$-direction),
see Ref.[\onlinecite{Lem03}] for a review.

There have been several mechanisms proposed for the driving force
of this phase transition:
the spin-phonon coupling like in a spin-Peierls system,\cite{Isobe96}
the electron-phonon coupling, the intra-ladder Coulomb repulsion
between $V^{4}$ and $V^{5}$-ions on the neighboring
rungs\cite{Most98,Thal98,Seo98,Most02},
or a combination of those interaction terms.\cite{Riera99}
The experiments seem however to rule out\cite{Most98,Most02,Lem03}
the original proposal\cite{Isobe96} for the spin-Peierls scenario,
while the zigzag charge ordering pattern predicted by the
Coulomb-driven scenarios of this phase
transition,\cite{Most98,Seo98,Most02} is now established
crytallographically.\cite{Smaalen02}
Another argument in favor of the key role played by the charge
ordering at this phase transition, comes from the fact that
the coupling between spins and zigzag-ordered charge
leads to\cite{Gros99} the observed dispersion\cite{Yosi98}
along the $a$-direction (perpendicular to the ladders) of the
gapped magnon mode in the low-temperature phase.

Since $\rm NaV_2O_5$ is an insulator, one can model the charge
degrees of freedom in $\rm NaV_2O_5$ on a given rung by a pseudospin
$\bm{\mathcal T}$. The system of quarter-filled coupled ladders can be
mapped onto a triangular lattice where one spin and one pseudospin reside
on the same site\cite{Most98,Deb00,Most02}, compare Fig.\ \ref{LatFig}.
Mostovoy and Khomskii in their study
of $\rm NaV_2O_5$ proposed the following mechanism for the
spin dimerization driven by charge ordering (i.e., spin gap generation
in the \textit{charge ordered phase}):\cite{Most98,Most02}
the antiferroelectric (AFE) in-ladder ordering (i.e., zigzag charge
ordering) is due to the inter-rung Coulomb repulsion. The ordering
pattern is shifted by a half of the inter-rung distance from ladder to
ladder (see Fig.\ref{LatFig}), and it
results in alternation of the spin-exchange coupling along
the ladder induced by its left and right neighbors. Then, as a
consequence of the spin-exchange dimerization, the systems acquires a
spin gap. Mostovoy and Khomskii\cite{Most98} also pointed out that
for this effect to occur,
it suffices for the dimerization term of the spin-exchange coupling
(i.e, the inter-ladder spin-pseudospin coupling) to
be linear over the charge order parameter. This mechanism resembles
so the familiar spin-Peierls scenario, with the proviso that
phonons are replaced by the charge order parameter.

Going deeper into this suggestion for the coupling which could be
a key to the problem of $\rm NaV_2O_5$, we perform a detailed
analysis of such a phase transition starting from a simplest possible
Hamiltonian compatible with symmetry restrictions.
The effective Hamiltonian is that of the Ising Model in Transverse Field
(IMTF) for pseudospins $\bm{\mathcal T}$, coupled to the spin-$\frac12$
$XY$-chains. Since the phase transition in $\rm NaV_2O_5$ is
\textit{thermal} in nature,
we treat the charge (pseudospin) sector of the Hamiltonian in the
molecular-field approximation, while retaining in our analysis the
experimentally observed 1D nature of the spin excitations.
Recent x-ray\cite{Ravy99,Gaulin00} and NMR\cite{Fagot00}
experiments showed large fluctuation regions on the both sides of
$T_c$ in $\rm NaV_2O_5$, and the order parameter critical index is
found to be $\beta \approx 0.17-0.19$. It is close to $\beta = 1/8$
of the 2D Ising model, indicating along with the correlation lengths
measurements\cite{Ravy99} on the 2D-Ising universality class of
this phase transition. According to Ref.[\onlinecite{Ravy99}] the
1D character of structural fluctuations manifests itself only
at $T \agt 60$K. This allows us to expect the mean-field treatment
to be a qualitatively reasonable approximation for the problem
we are dealing with.

We find that our spin-pseudospin model has a phase diagram with
two phases: the disordered one without charge order or
spin gap; and the ordered low-temperature phase with both the
anti-ferroelectric (zigzag) charge order and spin gap.
One of our most important findings is that the model
\textit{always} orders (continuously, via a second-order
transition) into that phase at low temperatures. The critical
temperature of this phase transition shows various regimes of
dependence on model's parameters, but it never vanishes as far
as the inter-ladder spin-pseudospin coupling is non-zero.
Physically, it means that the role of the inter-ladder coupling
is twofold: this coupling not only creates the spin gap when
the phase transition is driven by the charge (pseudospin)
ordering in the IMTF, but it also destroys the quantum
critical point of the IMTF and generates the charge-ordered and
spin-gapped phase in the case when charge
ordering in the pure IMTF is impossible.

The rest of the paper is organized as follows. In Section \ref{MoFo} we
define the model Hamiltonian and describe the methods employed to handle
it. In Section \ref{MF} we present the mean-field equations which govern
the critical behavior of the system. Section \ref{Tcr} presents the analysis
of the phase diagram, i.e., the critical temperature of the
anti-ferroelectric phase transition as a function of the Hamiltonian's
couplings. The results for the order parameter, spin and pseudospin
susceptibilities, specific heat capacity calculated in the vicinity of the
anti-ferroelectric phase transition are given in Section \ref{AFE}.
The summary of our results and the discussion on their application for
$\rm NaV_2O_5$ are presented in the final Section \ref{Concl}.
%
%%%%%%%%%%%%%%%%%%%%%%%%%%%%%%%%%%%%%%%%%%%%%%%%%%%%%%%%%%%%%%%%%%%%%%%%%%%%%%
\section{Model and Formalism}\label{MoFo}
%%%%%%%%%%%%%%%%%%%%%%%%%%%%%%%%%%%%%%%%%%%%%%%%%%%%%%%%%%%%%%%%%%%%%%%%%%%%%%
%
%
%%%%%%%%%%%%%%%%%%%%%%%%%%%%%%%%%%%%%%%%%%%%%%%%%%%%%%%%%%%%%%%%%%%%%%%%%%%%%%
\subsection{Model}\label{Model}
%%%%%%%%%%%%%%%%%%%%%%%%%%%%%%%%%%%%%%%%%%%%%%%%%%%%%%%%%%%%%%%%%%%%%%%%%%%%%%
%
We choose the following effective Hamiltonian:
\begin{eqnarray}
\label{Ham}
H= &-& \Omega \sum_{m,n} \mathcal T^z_{mn} +\frac12 J_{\textsc I} \sum_{m,n}
\mathcal T^x_{mn} \mathcal T^x_{m,n+1}
+ J_{\textsc H} \sum_{m,n} {\textbf S}_{mn} {\textbf S}_{m,n+1}
\\ \nonumber
&+&  \sum_{m,n} {\textbf S}_{mn}  {\textbf S}_{m,n+1}
\big[ J_{\textsc ST} \mathcal T^z_{mn} \mathcal T^z_{m,n+1}+
 \tilde \varepsilon
 \big( \mathcal  T^x_{m+1,n+1}- \mathcal T^x_{m-1,n}\big)
\big]
\end{eqnarray}
where the spin operators for a given site are defined as $S^i=\frac12 \sigma^i$
($i=x,y,z$) via the Pauli matrices $\sigma^i$. The pseudospin operators
are related in the same way to the Pauli matrices $\tau^i$:
$\mathcal T^i=\frac12 \tau^i$. 
We use the distinct notations for the Pauli matrices $\sigma, \tau$ in order
to emphasize the fact that they operate in different spaces (spin, pseudospin).
So the components, e.g.,  $\sigma^x$ and $\tau^x$ do not have the same
quantization axis. By labelling the two sets of Pauli's matrices
$\sigma, \tau$ with $x,y,z$ we mean only their conventional
representation and commutation relations. The subscript
$1 \leq m \leq \mathcal M$ stands for the number of ladder, while
$1 \leq n \leq \mathcal N$ numbers the rung.
We assume the Hamiltonian coupling constants to be positive.
Note in making comparison with the microscopic Hamiltonians derived for
$\rm NaV_2O_5$,\cite{Most98,Most02,Deb00} that in our Hamiltonian
(\ref{Ham}) we made the $\frac{\pi}{2}$-rotation
in the pseudospin space $\mathcal T^x \mapsto \mathcal T^z$,
$\mathcal T^z \mapsto -\mathcal T^x$. 

The first two terms on the right
hand side of Eq.(\ref{Ham}) correspond to the Ising Model in Transverse
Field (IMTF), the third term to the Heisenberg Hamiltonian, and in the last
one which describes the spin-pseudospin interaction
$\propto \tilde \varepsilon$,
we retain only linear term over the rung charge displacement operator
$\mathcal T^x$. This inter-ladder coupling term is allowed by the point
symmetry group $D_{2h}$ (the $\rm NaV_2O_5$ space group is
$D_{2h}^{13}-Pmmn$).\cite{Smo98}
Note that a similar intra-ladder spin-pseudospin coupling
$\sum_{mn}
{\textbf S}_{mn} {\textbf S}_{m,n+1}
 \big( \mathcal  T^x_{m,n+1}- \mathcal T^x_{m,n}\big)
$ would be odd with respect to the mirror-plane through the
bridging-oxygen (the $bc$-plane), therefore it is forbidden by
symmetry.

In the following we will work with the dimensionless Hamiltonian
${\mathcal H}=H/ \Omega$
\begin{eqnarray}
\label{Hamdls}
{\mathcal H}=&-& \sum_{m,n} \mathcal T^z_{mn} +
\frac12 g \sum_{m,n} \mathcal T^x_{mn} \mathcal T^x_{m,n+1}
\\ \nonumber
&+&  \sum_{m,n}D_{mn}
\big[ J +\lambda \mathcal T^z_{mn} \mathcal T^z_{m,n+1}+
\varepsilon \big( \mathcal T^x_{m+1,n+1}- \mathcal T^x_{m-1,n}\big)
\big]~,
\end{eqnarray}
dimensionless temperature $T  \rightarrow T/\Omega$,
and dimensionless couplings
\begin{eqnarray}
\label{Coupdls}
g=J_{\textsc I}/\Omega~,~~
J=J_{\textsc H}/\Omega~, \\ \nonumber
\lambda=J_{\textsc ST}/ \Omega~,~~
\varepsilon= \tilde \varepsilon / \Omega~,
\end{eqnarray}
where the dimerization operator
$D_{mn} \equiv {\textbf S}_{mn}\cdot{\textbf S}_{m,n+1}$.
For the applications we have in mind, the range of
the model's parameters under consideration will be restricted to the cases
\begin{equation}
\label{parange}
(J, \lambda) \alt  g, ~~ \varepsilon \ll (g, \textrm{max} \{J, \lambda \})
\end{equation}
In the present
study we consider the Hamiltonian with the $XY$ spin-spin
interaction, i.e.,
\begin{eqnarray}
\label{Sint}
D_{mn}=D_{mn}^{\textsc XY} &\equiv&
S_{mn}^x S_{m,n+1}^x+S_{mn}^y S_{m,n+1}^y \\ \nonumber
 &=& \frac12 \Big(
 S_{mn}^+ S_{m,n+1}^- +S_{mn}^- S_{m,n+1}^+
\Big)
\end{eqnarray}
where we used the conventional definition
$S^\pm \equiv S^x \pm i S^y$.
%
%%%%%%%%%%%%%%%%%%%%%%%%%%%%%%%%%%%%%%%%%%%%%%%%%%%%%%%%%%%%%%%%%%%%%%%%%%%%%%
\subsection{Molecular Field Approximation}\label{MFA}
%%%%%%%%%%%%%%%%%%%%%%%%%%%%%%%%%%%%%%%%%%%%%%%%%%%%%%%%%%%%%%%%%%%%%%%%%%%%%%
%
We proceed with the approximate treatment of our model in the following
way: The model Hamiltonian (\ref{Hamdls}) can be viewed as
$\mathcal H = \mathcal H_{\textrm{IMTF}} + \mathcal H_{\textrm{XY}}$.
$\mathcal H_{\textrm{IMTF}}$ is the Hamiltonian of the IMTF
in terms of the pseudospin operators $\mathcal T$
which is exactly solvable by its own. Dependence of the effective coupling
$J^{\textrm{eff}}_{mn}$ on $\mathcal T$ in the $XY$ spin Hamiltonian
$\mathcal H_{\textrm{XY}}= \sum J^{\textrm{eff}}_{mn} D_{mn}$
precludes the total Hamiltonian from being exactly solvable.

We apply a version of the Mean Field Approximation (MFA) which is
rather known as the Molecular Field method (Approximation) in the
theory of phase transitions\cite{Blinc74}. This approximation assumes
separability of the total Hamiltonian
density matrix $\rho$ over $S$ and $\mathcal T$, i.e.,
\begin{equation}
\label{STsep}
\rho=\rho^S \otimes \rho^{\mathcal T}
\end{equation}
and the following single particle (operator)
ansatz for the pseudospin density matrix
\begin{eqnarray}
\label{rhoT}
\rho^{\mathcal T} &=& \prod_{m,n}\rho^{\mathcal T}_{mn}, ~\\
\label{rhoTsing}
\rho^{\mathcal T}_{mn} &=& \frac{1}{Z_{mn}^{\mathcal T}}
\textrm{exp}(-\beta \textbf h_{mn} \bm{\mathcal T}_{mn})
\end{eqnarray}
Here $\textbf h_{mn}$ is the on-site vector of the molecular field to be
defined self-consistently, $\beta$ is inverse temperature (we set the Boltzmann
constant $k_B=1$),  and
\begin{equation}
\label{Zmn}
Z_{mn}^{\mathcal T}=2 \cosh \frac{\beta |\textbf h_{mn}|}{2}
\end{equation}
in the on-site partition function.

Since the independent spin-pseudospin averaging (\ref{STsep}) with the
single-particle density matrix (\ref{rhoTsing}) makes $J_{\textrm{eff}}$
a function of the average values of $\mathcal T$, no further approximation
is needed for the density matrix $\rho^S$.
The statistical mechanical problem with
\begin{equation}
\label{rhoS}
\rho^S = \frac{1}{Z_{\textrm{S}}} e^{-\beta \mathcal H_{\textrm{S}}}~,
\quad Z_{S}= \textrm{Tr} e^{-\beta \mathcal H_{\textrm{S}}},
\end{equation}
$ \mathcal H_{\textrm{S}}=\mathcal H_{\textrm{XY}}$ (here Tr includes
only spin-operator states) is exactly solvable,\cite{Lieb61} and will be
dealt with in the next subsection. To summarize more
qualitatively, we solve exactly the problem of independent spin $XY$
chains with their exchange constants determined by the molecular fields of
quasi-two-dimensional pseudospins.
%
%%%%%%%%%%%%%%%%%%%%%%%%%%%%%%%%%%%%%%%%%%%%%%%%%%%%%%%%%%%%%%%%%%%%%%%%%%%%%%
\subsection{Spin Hamiltonian}\label{Hxy}
%%%%%%%%%%%%%%%%%%%%%%%%%%%%%%%%%%%%%%%%%%%%%%%%%%%%%%%%%%%%%%%%%%%%%%%%%%%%%%
%
The standard Jordan-Wigner transformation\cite{Lieb61,Chak96} (JWT) when applied
independently to one-dimensional spin operators on a given ladder, results in
fermionic operators which commute on different ladders (see, e.g.,
Ref.[\onlinecite{Tsvelik95}]). Strictly speaking this does not cause
problems for our choice of Hamiltonian (\ref{Hamdls}), since its spin
part does not couple different ladders. However we find this situation
with the 1D JWT somehow unpleasant, especially keeping in mind possible
generalizations of the present Hamiltonian for future work. There are
generalizations of the JWT for $d>1$ available in the literature (see, e.g.,
Ref.[\onlinecite{GenJW,Azz93}]).

We will employ a version of the 2D JWT proposed by Azzouz\cite{Azz93},
particularly convenient for our model. We also include to
the spin Hamiltonian a term generated by the uniform external magnetic
field
\begin{equation}
\label{Hext}
\mathcal H_{\textrm ext} = \sum_{m,n} h_{M} S^z_{mn}
\end{equation}
The transformation is given by the following
equations:\cite{Azz93}
\begin{subequations}
\label{2DJW}
\begin{eqnarray}
S_{mn}^-  &=& \textrm{exp} \big(
-i \pi \sum_{k,l \in \mathfrak A_{mn}} n_{kl}
\big) c_{mn} \\
S_{mn}^+  &=& c_{mn}^\dagger
\textrm{exp} \big(
 i \pi \sum_{k,l \in \mathfrak A_{mn}} n_{kl}
\big)
\end{eqnarray}
\end{subequations}
where $n_{mn}=c_{mn}^\dagger c_{mn}$ is the fermionic number operator,
and the above summations include all rungs and ladders lying on the left
from the $m$-th ladder plus rungs lying below $n$-th rung at the $m$-th
ladder, i.e.,
\begin{equation}
\label{Amn}
k,l \in \mathfrak A_{mn} \Leftrightarrow
\{1 \leq k \leq m-1, 1 \leq l  \leq \mathcal N \} \cup
\{ k= m, 1 \leq l \leq n-1 \}
\end{equation}
The fermionic operators satisfy the canonical anticommutation relations
\begin{subequations}
\label{anticom}
\begin{eqnarray}
\{c_{mn}^\dagger, c_{kl} \} &=&\delta_{mk} \delta_{nl} \\
\{c_{mn}, c_{kl}         \} &=& 0
\end{eqnarray}
\end{subequations}
and
\begin{eqnarray}
\label{XYmap}
D_{mn}^{\textsc XY} &=& \frac12 (
c_{mn}^\dagger c_{m,n+1}+ c_{m,n+1}^\dagger c_{mn} ) \\
\label{Zmap}
S_{mn}^z &=& c_{mn}^\dagger c_{mn}- \frac12
\end{eqnarray}
For the effective coupling $J^{\textrm{eff}}_{mn}$ determined by the mean-field
values of the pseudospin operators we take a dimerized ansatz, so
\begin{equation}
\label{HxyMF}
\mathcal H_{\textrm{XY}}^{\textrm{MF}}=
\sum_{m,n} J^{\textrm{eff}}_{mn} D_{mn}^{\textsc XY}~, \quad
J^{\textrm{eff}}_{mn}= A_m+(-1)^nB_m
\end{equation}
Applying then the JWT defined above and a Bogolyubov
transformation,\cite{Lieb61} we obtain
\begin{equation}
\label{Hs}
\mathcal H_{\textrm{S}} \equiv
\mathcal H_{\textrm{XY}}^{\textrm{MF}} +\mathcal H_{\textrm ext}
=- \frac12 \mathcal{MN} h_{M} +\sum_{m,q,\nu} E_{m \nu}(q)
d_{mq \nu}^\dagger d_{mq \nu}
\end{equation}
where the new fermionic operators $d_{mq \nu}$ also satisfy the canonical
anticommutation relations, and their spectrum is
\begin{equation}
\label{E}
E_{m \nu}(q)=h_{M}+\nu \sqrt{ A_m^2 \cos^2qa+ B_m^2 \sin^2qa}
\end{equation}
The extra index $\nu = \pm 1$ is due to dimerization, i.e.,
doubling of fermion species, summation over $q$ includes the
reduced Brillouin zone $[-\frac{\pi}{2a},\frac{\pi}{2a}]$,
and $a$ is the distance between rungs (we will set $a=1$ in
the following). For more details, see, e.g.,
Refs.[\onlinecite{Heeger88,Holi01}]. The operator of the
total $z$-component of spin per rung reads
\begin{equation}
\label{sz}
s_z \equiv \frac{1}{\mathcal{MN}}\sum_{m,n}S_{mn}^z=
- \frac12 +
\frac{1}{\mathcal{MN}} \sum_{m,q,\nu} d_{mq \nu}^\dagger d_{mq \nu}
\end{equation}
The spin contribution to
the free energy per
rung from the Hamiltonian $\mathcal H_{\textrm{S}}$ (\ref{Hs})
[cf. Eqs.(\ref{rhoS})] is
\begin{eqnarray}
\label{fs}
f_{\textrm{S}} &\equiv&  - \frac{1}{\beta \mathcal{MN}}
\ln Z_{\textrm{S}} \\ \nonumber
 &=& - \frac{\ln 2}{\beta}- \frac{1}{\pi \beta \mathcal M}
\sum_{m,\nu} \int_0^{\frac{\pi}{2}}
dq \ln \cosh \frac{\beta E_{m \nu}(q)}{2}
\end{eqnarray}
%
%
%%%%%%%%%%%%%%%%%%%%%%%%%%%%%%%%%%%%%%%%%%%%%%%%%%%%%%%%%%%%%%%%%%%%%%%%%%%%%%
\section{Mean-Field Equations }\label{MF}
%%%%%%%%%%%%%%%%%%%%%%%%%%%%%%%%%%%%%%%%%%%%%%%%%%%%%%%%%%%%%%%%%%%%%%%%%%%%%%
The molecular fields are defined from the condition that they
minimize the mean-field free
energy $\mathcal F =\langle \mathcal H \rangle+
T \langle \ln \rho \rangle$.
The angular brackets stand for
averaging with the mean-field density matrix (\ref{STsep}), and the
Hamiltonian $\mathcal H$ is given by Eq.(\ref{Hamdls}). We consider
the case when the external magnetic field $h_M$ is absent. (Note
that contrary to $f_{\textrm{S}}$ (\ref{fs}), the quantity
$f_{\mathcal T}$ defined in the same fashion via $Z^{\mathcal T}$ (\ref{Zmn})
is not the pseudospin free energy).

We take the following ansatze for the Ising pseudospin magnetizations
(i.e., the charge ordering parameters in terms of the real physical
quantities)
\begin{eqnarray}
\label{mz}
\langle \mathcal T_{mn}^z \rangle &=& m_z \\
\label{mx}
\langle \mathcal T_{mn}^x \rangle &=&(-1)^{m+n}m_x
\end{eqnarray}
So, similar to the case of the IMTF ($g>0$) we assume the possibility
of the anti-ferroelectric (AFE) in-ladder charge ordering (i.e., the
zigzag ordering\cite{Most98}), alternating however from ladder to ladder.
It is easy to see from the Hamiltonian (\ref{Hamdls}) that ansatz
(\ref{mx}) creates a dimerization in the spin sector, so a natural
assumption for the dimerization operator average is
\begin{equation}
\label{Dmn}
\langle D_{mn} \rangle =-[t+(-1)^{m+n} \delta]
\end{equation}
Then the average mean-field energy per rung in the thermodynamic limit
$\mathcal{MN} \to \infty$ is
\begin{equation}
\label{Emf}
{\mathcal E}_{\textrm{MF}}
 \equiv \frac{\langle \mathcal H \rangle}{\mathcal{MN}}=
-m_z-\frac12 g m_x^2
-\big( Jt+\lambda t m_z^2+2 \varepsilon \delta m_x \big)
\end{equation}
The molecular Weiss fields $h_{mn}^z=-h_z$; $h_{mn}^x=(-1)^{m+n+1}h_x$;
$J^{\textrm{eff}}_{mn}=a+(-1)^{m+n}b$ are defined by the following
equations:
\begin{subequations}
\label{MolF}
\begin{eqnarray}
h_z &=&- \frac{\partial \mathcal E_{\textrm{MF}}}{\partial m_z}
= 1+ 2\lambda t m_z \\
h_x &=&- \frac{\partial \mathcal E_{\textrm{MF}}}{\partial m_x}=
g m_x+ 2\varepsilon \delta \\
a   &=&- \frac{\partial \mathcal E_{\textrm{MF}}}{\partial t}=
 J+\lambda m_z^2 \\
b   &=&- \frac{\partial \mathcal E_{\textrm{MF}}}{\partial \delta}=
2 \varepsilon m_x
\end{eqnarray}
\end{subequations}
The order parameters (\ref{mz},\ref{mx},\ref{Dmn}) obtained as
partial derivatives of the corresponding quantities
$f_S,f_{\mathcal T}$ with
respect to their conjugate Weiss fields (\ref{MolF}), should
be determined from the system of four coupled equations
\begin{subequations}
\label{EqMF}
\begin{eqnarray}
m_z
&=&\frac12
\frac{1+2\lambda t m_z}{\mathfrak h} \tanh\frac{\beta \mathfrak h}{2} \\
m_x
&=&\frac{m_x}{2}
\frac{g+2\varepsilon \eta}{\mathfrak h} \tanh\frac{\beta \mathfrak h}{2} \\
t &=& \frac{1}{\pi} \int_0^{\frac{\pi}{2}}
d \varphi \frac{\cos^2 \varphi}{\xi(\varphi)}
\tanh  \tilde \beta \xi(\varphi)
\equiv \frac{1}{\pi}t_n(\Delta, \tilde \beta) \\
\eta &=& \frac{\Delta}{\pi m_x} \int_0^{\frac{\pi}{2}}
d \varphi \frac{\sin^2 \varphi}{\xi(\varphi)}
\tanh \tilde \beta \xi(\varphi)
\equiv \frac{\Delta}{\pi m_x} \eta_n(\Delta, \tilde \beta)
\end{eqnarray}
\end{subequations}
where $\mathfrak h =\sqrt{h_x^2+h_z^2}$ is the absolute value of the Ising
molecular field, and instead of $\delta$ we use the new parameter $\eta$
\begin{equation}
\label{eta}
\delta \equiv m_x \eta
\end{equation}
We also introduced the auxiliary parameters
\begin{subequations}
\label{Param}
\begin{eqnarray}
\xi(\varphi) &\equiv& \sqrt{ \cos^2 \varphi+\Delta^2 \sin^2 \varphi  } \\
\Delta       &\equiv& \frac{2 \varepsilon m_x}{J+\lambda m_z^2} \\
\tilde \beta   &\equiv& \frac{\beta}{2}(J+\lambda m_z^2)
\end{eqnarray}
\end{subequations}
The following useful relationship holds at any temperature
\begin{equation}
\label{mabs}
m_x^2+m_z^2=\frac14 \tanh^2 \frac{\beta \mathfrak h}{2}
\end{equation}

Alternatively, the mean-field equations (\ref{EqMF}) can be derived from
the free energy (per rung, $h_M=0$)
\begin{eqnarray}
\label{FE}
f &=& 2 \lambda t m_z^2 +\frac12 g m_x^2+
2\varepsilon \delta m_x -T\ln4\\ \nonumber
  &-& T \Big(
\ln \cosh\frac{\beta \mathfrak h}{2}
+\frac{2}{\pi} \int_0^{\frac{\pi}{2}}
d \varphi \ln \cosh \tilde \beta \xi(\varphi) \Big)~,
\end{eqnarray}
understood as $f(m_z,m_x,t,\delta)$, by minimization
over its variables. [$ \mathfrak h$ is
defined according to Eqs.(\ref{MolF}).] Thus, $f$ plays a role of the Landau
functional, defining parameters $m_z,m_x,t,\delta$.

Before presenting the analysis of Eqs.(\ref{EqMF}), let us describe
their main properties more qualitatively. The system of these four
coupled equations determines four unknown mean-field parameters
$m_z, m_x, t, \eta~(\delta)$ as functions of the effective Hamiltonian
couplings and temperature. The first couple of Eqs.(\ref{EqMF})
in the case $\lambda=\varepsilon=0$ is familiar from the mean-field
treatment of the pure IMTF (see, e.g, Ref.[\onlinecite{Blinc74}]),
and has some similar properties with the pure case even for non-zero
$\lambda$ and $\varepsilon$.
There is always a non-trivial solution  $0 < m_z \leq \frac12$ for
the field-induced magnetization
(the lower bound value $m_z=0$ is attained in the limit $g \to \infty$).
Below certain critical temperature $T_c$  a non-trivial
solution $m_x \neq 0$ appears. $m_z$ varies continuously across $T_c$,
contrary to the case of the IMTF ($\lambda=\varepsilon=0$) when it stays
constant ($=1/g$) for all $T <T_c$. Non-zero $m_x$ results in generation of
the alternating term $b$ in the exchange interaction
$J^{\textrm{eff}}_{mn}$ and the non-zero dimerization parameter $\delta$.
Consequence of $b \neq 0$ is a spin gap.
Note that in the disordered phase ($m_x=0$),
Eq.(\ref{EqMF}d) is an empty statement at $T \neq 0$, since  $\delta =0$
as well. The parameter $t$ is non-critical: it is some continuous
non-negative function.

At $T \leq T_c$ one equation from the pair (\ref{EqMF}a,b) can be written
in the form
\begin{equation}
\label{mzbe}
m_z^{-1} = g+ \frac{4 \varepsilon^2}{\pi(J+\lambda m_z^2)} \eta_n
 -\frac{2\lambda }{\pi} t_n , ~~T \leq T_c
\end{equation}
more convenient for further analyses.
%
%
%%%%%%%%%%%%%%%%%%%%%%%%%%%%%%%%%%%%%%%%%%%%%%%%%%%%%%%%%%%%%%%%%%%%%%%%%%%%%%
\section{Critical Temperature}\label{Tcr}
%%%%%%%%%%%%%%%%%%%%%%%%%%%%%%%%%%%%%%%%%%%%%%%%%%%%%%%%%%%%%%%%%%%%%%%%%%%%%%
%
%
At $T=T_c$ we have Eqs.(\ref{EqMF}a) as
\begin{equation}
\label{mzTc}
m_z = \frac12 \tanh\frac{\beta_c }{2}
\big(1+\frac{2\lambda m_z}{\pi}t_n \big)~,
\end{equation}
another equation for $m_z$ (\ref{mzbe}), and
parameters $t_n,\eta_n$ are given by Eqs.(\ref{EqMF}c,d) with $\Delta=0$.
The latter two functions have the following expansions:
\begin{equation}
\label{tnas}
t_n(0,x) \approx  \left\{
                \begin{array}{ll}
 \frac{\pi}{4} x(1-\frac14 x^2) + \mathcal O (x^5 ) , & x <1\\
  1-\frac{\pi^2}{24} \frac{1}{x^2}, & x >1
                \end{array}
       \right.
\end{equation}
and
\begin{equation}
\label{etanas}
\eta_n(0,x) \approx \left\{
                \begin{array}{ll}
  \frac{\pi}{4} x(1-\frac{1}{12}x^2) + \mathcal O(x^5 ) , & x <1 \\
 \ln \mathbb A x+ \mathcal O(\frac{1}{x^2}), & x >1
                \end{array}
       \right.
\end{equation}
where $\mathbb A \equiv\frac{8}{\pi \textrm{e}^{1- \gamma}} \approx 1.6685$,
and\ $\gamma \approx 0.5772$ is Euler's constant.
%
%%%%%%%%%%%%%%%%%%%%%%%%%%%%%%%%%%%%%%%%%%%%%%%%%%%%%%%%%%%%%%%%%%%%%%%%%%%%%%
\subsection{Case $\varepsilon=0$: re-entrance}\label{eps0}
%%%%%%%%%%%%%%%%%%%%%%%%%%%%%%%%%%%%%%%%%%%%%%%%%%%%%%%%%%%%%%%%%%%%%%%%%%%%%%
%
In order to better understand the model, let us start with the analysis of the
mean-field equations for the case when the dimerization coupling
is absent, i.e, $\varepsilon=0$. Then from Eqs.(\ref{mzbe},\ref{mzTc}) we
find the critical temperature $T_c \equiv 1/\beta_c$
\begin{equation}
\label{Tceps0}
T_c=\frac{m_zg}{\ln \frac{1+2m_z}{1-2m_z}}
\end{equation}
where $m_z$ is the solution of the following equation
\begin{equation}
\label{mzeps0}
m_z^{-1}=g-\frac{2 \lambda}{\pi}
t_n \bigg(0, \frac{J+\lambda m_z^2}{2m_zg}\ln \frac{1+2m_z}{1-2m_z}
  \bigg).
\end{equation}
It can be shown from Eqs.(\ref{mzbe},\ref{mzTc},\ref{tnas}) that
for large Ising coupling $g\gg g_{\lambda}$ where
\begin{equation}
\label{glam}
g_{\lambda} \equiv 2 \Big(1+ \frac{\lambda}{\pi} \Big),
\end{equation}
the critical temperature $ T_c \approx g/4$ is large , while
$m_z \approx 1/g$ is small. So, the behavior of these parameters is the
same as in the pure IMTF.\cite{Blinc74}
It can also be shown that at $g=g_{\lambda}$ the critical temperature
$T_c$ becomes zero ($m_z=\frac12$). Thus $g_{\lambda}$ plays a role of the
renormalized (due to the coupling $\lambda$) mean-field critical point
$g^*=g_{\lambda}(\lambda=0)=2$ of the pure IMTF, where $T_c$ vanishes
and no ordering in
$m_x$ occurs for $g <g^*$.\cite{Blinc74} (The mean-field $g^*$
is two times less than the value known for the exactly
solvable 1D IMTF. For 2D and 3D IMTF the values of $g^*$ are known only
from approximations and/or numerics, but they are bigger
then the mean-field predictions.\cite{Chak96})

However, the role of $\lambda$ is more subtle than a simple
shift in the parameters comparatively to the pure IMTF.
The critical line $T_c(g)$ manifests re-entrant behavior in
the region $g \alt g_{\lambda}$ or, to state it differently,
the function $g(T_c)$ is non-monotonic, contrary to the case
$\lambda =0$. (See Fig.\ \ref{TcFig}). Also, $T_c(g_{\lambda})=0$
does not mean that  $g_{\lambda}$ is the minimal coupling
$g_{\textrm{min}}$ at which $m_x$-ordering can occur.
How pronounced the re-entrance is, i.e., the width
of the re-entrance region
\begin{equation}
\label{reentr}
g_{\textrm{min}} <g<g_{\lambda},
\end{equation}
depends
on the relative values of model's couplings $J,\lambda, g$. A
more detailed analysis of the re-entrance in
this model will be presented in a separate paper.\cite{CGlet03}
At this time, we say that any value of the coupling $\lambda \neq 0$
always generates the re-entrance on the phase diagram $g(T_c)$,
i.e., $g_{\textrm{min}} \to g_{\lambda} \to 2$ only when
$\lambda \to 0$.
Characteristic results of
the numerical solution of our equations are given in Fig.\ \ref{TcFig}.
At $J \ll \lambda$ this feature is pronounced most [the
case $J=0, \lambda=1$ shown in this figure], while at $J \sim \lambda$
it is more smeared [see the curve at $J=\lambda=1$].
Notice that when $g \to g_{\lambda}$,  the curve $T_c(g)$ approaches zero
normally to the $g$-axis.

More qualitatively, the origin of the re-entrance
can be understood from the free energy (\ref{FE}) which is
minimized by our mean-filed equations.
First two terms on the r.h.s. of Eq.(\ref{FE}) augment the
free energy $f$ with the increase of $m_z$ and/or $m_x$, while
at the same time the last two terms on the r.h.s. of that equation,
explicitly proportional to the temperature, decrease $f$ via
the parameters $\mathfrak h, \tilde \beta, \Delta$. The interplay
of this contributions to $f$ involves, apart of the temperature,
the couplings $g, J, \lambda$. The latter possess a range
[the re-entrance region (\ref{reentr})] wherein the minimal free
energy is achieved by re-distribution of the values of
$m_z(T)$ and $m_x(T)$. In particular, within this region, upon say,
decreasing $T$, the system goes smoothly from the disordered phase
$m_x(T)=0$ to the ordered one $m_x(T) \neq 0$, and back to
$m_x(T)=0$. Note also from Eq.(\ref{mzTc}), that $\lambda$ creates a
self-consistent renormalization of the external field [$\Omega$,
in terms of the dimensionful Hamiltonian (\ref{Ham})], resulting
in, particularly,  the temperature dependence of $m_z$ in the ordered
phase. (Compare to the pure IMTF where at $T<T_c$ it is frozen
at $m_z=1/g$). According to Eq.(\ref{tnas}) the effective external field
$1+2 \lambda t_n m_z/ \pi$ grows at low temperature,
bringing $m_z$ closer to its maximal value $1/2$ and in the same time,
forcing $m_x$ to decrease [cf. Eq.(\ref{mabs})]. This is another
argument in the way to qualitatively explain why the system can re-enter
the disordered phase upon decreasing the temperature. However, we should
point out that these results and arguments are based on the mean-field
analysis, so whether or not the re-entrance survives a more sophisticated
treatment of the model, is not clear at this time.
%
%%%%%%%%%%%%%%%%%%%%%%%%%%%%%%%%%%%%%%%%%%%%%%%%%%%%%%%%%%%%%%%%%%%%%%%%%%%%%%
\subsection{Case $\varepsilon \neq 0$: double re-entrance}\label{epsnon0}
%%%%%%%%%%%%%%%%%%%%%%%%%%%%%%%%%%%%%%%%%%%%%%%%%%%%%%%%%%%%%%%%%%%%%%%%%%%%%%
%
An important qualitative change in the critical behavior of the model
occurs upon switching on the dimerization coupling $\varepsilon$.
At any $\varepsilon \neq 0$ the critical temperature is never zero,
even if $g \ll g_{\lambda}$.
Let us note first that when $\{ \Delta, T \} \to 0$ then
$\eta_n \sim -\ln (\textrm{min}\{\Delta,T \})$ is divergent, so this limit
should be treated with care. When $g \gg g_{\lambda}$ [cf. also conditions
(\ref{parange})] the behavior of the critical temperature and magnetization
$m_z$ is the same as in the case $\varepsilon =0$.
Upon decreasing $g$, the critical temperature $T_c$ decreases,
and according to Eq.(\ref{mzTc})
$m_z \approx \frac12$, while from Eqs.(\ref{mzbe},\ref{tnas},\ref{etanas})
$m_z^{-1} \approx c_1 -c_2 \cdot\ln T_c$, where $c_1,c_2$ are some constants.
From the consistency of those equations we conclude that for any finite
$\varepsilon$ the critical temperature never vanishes. To put it
differently, coupling $\varepsilon$ destroys the quantum critical
point of the IMTF. This constitutes a very important feedback effect of the
spin chains on the charge degrees of freedom.

Characteristic numerical results for $T_c(g)$ are shown in Fig.\ \ref{TcFig}.
Similar to the case $\varepsilon =0$, the critical temperature shows the
re-entrance in the region $g \alt g_{\lambda}$. Moreover, instead of going to
zero at $g = g_{\lambda}$ as for $\varepsilon =0$, the curve $T_c(g)$ turns
left, and $T_c$ remains finite even at $g=0$, albeit exponentially small in
$\varepsilon$. For the two curves discussed above in the case
$\varepsilon =0$, we show in Fig.\ \ref{TcFig} their counterparts at
$\varepsilon =0.1$. As one can see from the whole curve $T_c(g)$, at
bigger values of $\lambda$ the system manifests a well-pronounced
double re-entrance.

It follows from Eqs(\ref{mzbe},\ref{mzTc},\ref{tnas},\ref{etanas}) that
the left turn of the critical temperature and its failure to vanish
at $g=g_{\lambda}$ can be analytically
described as a BCS-type solution for $T_c(g)$, generated by finite
$\varepsilon$.
Approximate solutions for $T_c(g)$ found in two regions of $g$
are
\begin{equation}
\label{Tcas}
 T_c \approx  \left\{
                \begin{array}{ll}
  \frac{g}{4}, & g \gg g_{\lambda} \\[0.2cm]
 \frac{\mathbb A \tilde J }{2} \textrm{exp}
 \big[- \frac{\pi \tilde J}{4 \varepsilon^2}(g_{\lambda}-g) \big]
  , &\textrm{BCS regime}
                \end{array}
       \right.
\end{equation}
where
\begin{equation}
\label{Jbar}
\tilde J \equiv J +\frac{\lambda}{4}
\end{equation}
The boundary where the low-temperature BCS regime sets in and the related
formulas are applicable, is given by the condition
\begin{equation}
\label{BCScond}
\textrm{BCS regime}:~
g < g_{\lambda} +\frac{4 \varepsilon^2}{\pi \tilde J}
\end{equation}
The field-induced magnetization $m_z$ in these regions is
\begin{equation}
\label{mzas}
 m_z \approx  \left\{
                \begin{array}{ll}
  \frac{1}{g}, & g \gg g_{\lambda} \\[0.2cm]
  \frac12 , & \textrm{BCS regime}
                \end{array}
       \right.
\end{equation}
%
%%%%%%%%%%%%%%%%%%%%%%%%%%%%%%%%%%%%%%%%%%%%%%%%%%%%%%%%%%%%%%%%%%%%%%%%%%%%%%
\subsection{Case $\varepsilon \neq 0$, $\lambda=0$: no
re-entrance}\label{epsnon0lambda0}
%%%%%%%%%%%%%%%%%%%%%%%%%%%%%%%%%%%%%%%%%%%%%%%%%%%%%%%%%%%%%%%%%%%%%%%%%%%%%%
%
The analytical treatment in the intermediate regime when
$g \sim g_{\lambda}$ is involved due to re-entrance. The
situation simplifies when $\lambda =0$ and the re-entrance is absent
(see the curve for $J=1,\lambda=0, \varepsilon=0.1$ in Fig.\ \ref{TcFig}).
Then both the intermediate $g \agt g^*$ ($g^*=2$) and
the BCS regimes are well described by the approximate
equation
\begin{equation}
\label{eqapr}
g-g^* +\frac{4\varepsilon^2}{\pi J} \ln \frac{\mathbb A J}{2 T_c}
=4 e^{-1/T_c}
\end{equation}
In particular, at the IMTF quantum critical point
on finds that it is destroyed by
the inter-ladder dimerization coupling $\varepsilon$, resulting in
\begin{equation}
\label{TcCP}
T_c \approx \ln ^{-1} \Big(\frac{\pi J}{\varepsilon^2} \Big),~g=g^*
\end{equation}
So when the re-entrance is absent, upon decreasing $g$ the critical
temperature $T_c$ monotonically decreases from the IMTF linear regime
$T_c \propto g$ via the logarithmic dependence (\ref{TcCP}) towards
the exponential BCS regime (\ref{Tcas}). It is
interesting to note that a similar inverse-log dependence
of the transition temperature near quantum criticality has been found
in a recent molecular-field study of $\rm Cu_2Te_2O_5Br_2$,
a spin-system containing coupled spin-tetrahedra \cite{Gros03}.
%
%
%%%%%%%%%%%%%%%%%%%%%%%%%%%%%%%%%%%%%%%%%%%%%%%%%%%%%%%%%%%%%%%%%%%%%%%%%%%%%%
\section{Properties of the AFE Phase}\label{AFE}
%%%%%%%%%%%%%%%%%%%%%%%%%%%%%%%%%%%%%%%%%%%%%%%%%%%%%%%%%%%%%%%%%%%%%%%%%%%%%%
%
%%%%%%%%%%%%%%%%%%%%%%%%%%%%%%%%%%%%%%%%%%%%%%%%%%%%%%%%%%%%%%%%%%%%%%%%%%%%%%
\subsection{AFE order parameter}\label{OrdPar}
%%%%%%%%%%%%%%%%%%%%%%%%%%%%%%%%%%%%%%%%%%%%%%%%%%%%%%%%%%%%%%%%%%%%%%%%%%%%%%
%
In the ordered phase we can derive from Eqs.(\ref{MolF},\ref{EqMF},\ref{mabs})
the following equation
\begin{equation}
\label{EqSt}
\frac12 \frac{1}{\sqrt{m_x^2+m_z^2}}
\tanh \Big( \frac{\beta g_{\varepsilon}}{2}\sqrt{m_x^2+m_z^2} \Big)
=1
\end{equation}
where
\begin{equation}
\label{geps}
g_{\varepsilon} \equiv
g+ \frac{4 \varepsilon^2}{\pi(J+\lambda m_z^2)} \eta_n
\end{equation}
To determine the temperature behavior of the AFE order parameter $m_x$
in the immediate vicinity of the critical temperature where
\begin{equation}
\label{tau}
 \tau \equiv  \frac{T_c-T}{T_c} \ll 1~,
\end{equation}
we need to expand Eq.(\ref{EqSt}) near $T_c$, taking into account that all
parameters $m_x,m_z,g_{\varepsilon}$ entering this equation are
temperature-dependent and related via the mean-field equations.

It can be shown that to leading order
\begin{eqnarray}
\label{tnEx}
t_n(\Delta, \tilde \beta) &\approx& t_n(0, \tilde \beta)
+\Delta^2 \dot{t}_{n,\Delta^2} \\
\label{etanEx}
\eta_n(\Delta, \tilde \beta) &\approx& \eta_n(0, \tilde \beta)
+\Delta^2 \dot{\eta}_{n,\Delta^2}
\end{eqnarray}
with the partial derivatives given by the following equations
\begin{eqnarray}
\label{tnD}
\dot{t}_{n,\Delta^2} &=&
\frac{\partial t_n(\Delta,\tilde \beta)}{\partial \Delta^2}
\Big \vert_{\Delta=0}=-\Big[ a(\tilde \beta)-t_n(0, \tilde \beta)
\Big]\\
\label{etanD}
\dot{\eta}_{n,\Delta^2} &=&
\frac{\partial \eta_n(\Delta,\tilde \beta)}{\partial \Delta^2}
\Big \vert_{\Delta=0}=-\Big[ \frac12 a(\tilde \beta)+
\frac14 b(\tilde \beta) -\eta_n(0, \tilde \beta) \Big]
\end{eqnarray}
and
\begin{subequations}
\label{ab}
\begin{eqnarray}
a(\beta) &\equiv& \frac12 \int_0^1 \frac{\beta dz}{\cosh^2 \beta z}
\ln \frac{1+\sqrt{1-z^2}}{z} \\
b(\beta) &\equiv& \int_0^1 \frac{\beta dz}{\cosh^2 \beta z}
\bigg[ \frac{1}{\sqrt{1-z^2}}+2\beta\sqrt{1-z^2} \frac{\tanh\beta z}{z}
\bigg]
\end{eqnarray}
\end{subequations}
Using the above equations in the leading-order expansion of
Eqs.(\ref{mzbe},\ref{EqSt}) near $T_c$, we can show that
\begin{equation}
\label{SecOrd}
 m_x^2 \propto \tau
\end{equation}
along the whole line of the critical temperature depicted in Fig.\ \ref{TcFig},
i.e., $T_c$ defines a second-order phase transition, and the order parameter
has the mean-field critical index $1/2$. In general, the coefficient of
proportionality in the above relationship has to be defined numerically.
For the regimes of strong Ising coupling and BCS, the
order parameter can be calculated analytically.

For the functions defined by Eqs.(\ref{ab}) the following expansions are
obtained:
\begin{equation}
\label{aas}
a(x) \approx  \left\{
                \begin{array}{ll}
 \frac{\pi}{4} x(1-\frac16 x^2) + \mathcal O (x^5 ) , & x <1\\[0.2cm]
  \frac12 \ln \mathbb A e x +  \mathcal O (\frac{1}{x^2}), & x >1
                \end{array}
       \right.
\end{equation}
and
\begin{equation}
\label{bas}
b(x) \approx \left\{
                \begin{array}{ll}
  \frac{\pi}{2} x(1+\frac12 x^2) + \mathcal O(x^5 ) , & x <1 \\[0.2cm]
  \frac12 +\frac{14 \zeta(3)}{\pi^2}x^2 +
  \mathcal O(\frac{1}{x^2}), & x >1
                \end{array}
       \right.
\end{equation}
where $\mathbb A$ is defined below Eq.(\ref{etanas}), $e$ is the exponential
constant and $\zeta(x)$ - Riemann's
zeta-function. Combining these asymptotics with those given by
Eqs.(\ref{tnas},\ref{etanas}), we find
\begin{equation}
\label{mxExpas}
 m_x^2 \approx  \left\{
                \begin{array}{ll}
 \frac{g^2-4}{2g^2} \tau
  , & g \gg g_{\lambda} \\[0.2cm]
 \frac{2 \pi^2}{7 \zeta(3)} \frac{T_c^2}{\varepsilon^2} \tau
  , &~\textrm{BCS regime}
                \end{array}
       \right.
\end{equation}
Notice that $m_z$ is continuous across $T_c$, however as follows from
Eqs.(\ref{mzbe},\ref{tnEx},\ref{etanEx}) in the AFE phase
$m_z(T,m_x)=m_z(T,0)+\mathcal O(m_x^2)$, so its
derivative with respect to  temperature has a finite
discontinuity at $T_c$.
%
%%%%%%%%%%%%%%%%%%%%%%%%%%%%%%%%%%%%%%%%%%%%%%%%%%%%%%%%%%%%%%%%%%%%%%%%%%%%%%
\subsection{$T=0$}\label{T0}
%%%%%%%%%%%%%%%%%%%%%%%%%%%%%%%%%%%%%%%%%%%%%%%%%%%%%%%%%%%%%%%%%%%%%%%%%%%%%%
%
At zero temperature the parameters $t_n$ and $\eta_n$ are given in terms of the
complete elliptic integral of the first and second kind, such that
\begin{eqnarray}
\label{tn0}
t_n &=& \frac{E(1-\Delta^2)-\Delta^2K(1-\Delta^2)}{1-\Delta^2}       \\
\label{etan0}
\eta_n &=& \frac{K(1-\Delta^2)-E(1-\Delta^2)}{1-\Delta^2}
\end{eqnarray}
According to Eq.(\ref{mabs}) at $T=0$: $m_x^2+m_z^2=\frac14$, so we can
establish the range within which $\Delta$ can lie [cf. Eq.(\ref{Param}b)]
\begin{equation}
\label{DeltaRange}
0 \leq m_x \leq \frac12 \quad \Longleftrightarrow \quad
0 \leq \Delta \leq \frac{\varepsilon}{J} \ll 1
\end{equation}
where the last strong inequality is just the condition for Hamiltonian's
couplings we are implying from physical grounds. Note that
$\Delta_{\textrm{max}} \equiv \varepsilon / J $
is the absolute maximum $\Delta$ can reach.
Then we can safely make
expansions up to terms $\mathcal O (\Delta^2 \ln \Delta)$
\begin{eqnarray}
\label{tn0as}
t_n &\approx& 1          \\
\label{etan0as}
\eta_n &\approx& \ln \frac{4}{e \Delta}
\end{eqnarray}
which can be used in Eq.(\ref{mzbe}) for any $g$.
At large Ising couplings $g \gg g_{\lambda}$, $m_z$ is small, while
$m_x$ is close to its maximal value ($\Delta \approx 2 \varepsilon m_x/J$):
\begin{eqnarray}
\label{mz0P}
m_z^{-1}  &\approx& g- \frac{2\lambda}{\pi} +\frac{4\varepsilon^2}{\pi J}
\ln \frac{4J}{e \varepsilon}          \\
\label{mx0P}
m_x &\approx& \frac12 -m_z^2
\end{eqnarray}
For smaller couplings $g \leq g_{\lambda}$ the order parameter $m_x$ is small
($\Delta \approx \frac{2 \varepsilon m_x}{\tilde J}$):
\begin{equation}
\label{mx0asy}
 m_x \approx  \left\{
                \begin{array}{ll}
 \frac{\varepsilon}{  \sqrt{2 \pi \tilde J} }
\ln ^{\frac12} \Big(\frac{8 \pi \tilde J^3 }{e^2 \varepsilon^4 }
 \Big),
 & g = g_{\lambda} \\[0.3cm]
 \frac{2 \tilde J}{ e \varepsilon} \textrm{exp}
 \big[- \frac{\pi \tilde J}{4 \varepsilon^2}(g_{\lambda}-g) \big]
  , &\textrm{BCS regime}
                \end{array}
       \right.
\end{equation}
while
\begin{equation}
\label{mz0asy}
m_z \approx \frac12 -m_x^2
\end{equation}
is close to its maximal value. The dimerization parameter $\delta$
to leading order reads
\begin{equation}
\label{delta0asy}
 \delta \approx  \left\{
                \begin{array}{ll}
\big( \frac12 -\frac{1}{g^2} \big) \ln \frac{4J}{e \varepsilon},
& g \gg g_{\lambda} \\[0.3cm]
\frac{\varepsilon}{\sqrt{2 \pi \tilde J} }

\ln ^{\frac32} \Big(\frac{8 \pi \tilde J^3 }{e^2 \varepsilon^4 }
 \Big),
 & g = g_{\lambda} \\[0.3cm]
\frac{\pi \tilde J^2  (g_{\lambda}-g)  }{2 e \varepsilon^3} \textrm{exp}
 \big[- \frac{\pi \tilde J }{4 \varepsilon^2}(g_{\lambda}-g) \big]
  , &\textrm{BCS regime}
                \end{array}
       \right.
\end{equation}
%
%%%%%%%%%%%%%%%%%%%%%%%%%%%%%%%%%%%%%%%%%%%%%%%%%%%%%%%%%%%%%%%%%%%%%%%%%%%%%%
\subsection{Spin susceptibility}\label{chiSpin}
%%%%%%%%%%%%%%%%%%%%%%%%%%%%%%%%%%%%%%%%%%%%%%%%%%%%%%%%%%%%%%%%%%%%%%%%%%%%%%
%
The zero-field spin susceptibility per rung (i.e., per spin)
is given by
\begin{equation}
\label{chiS}
\chi_s \equiv -\frac{\partial \langle s_z \rangle}{ \partial h_M}
\Big \vert_{h_M=0}
=-\frac{\partial^2 f_{\textrm{S}} }{ \partial h_M^2}
\Big \vert_{h_M=0} =
\frac{\beta}{2 \pi} \int_0^{\frac{\pi}{2}}
\frac{d \varphi}{\cosh^2 \tilde \beta \xi(\varphi) }
\end{equation}
Its temperature dependence is similar to that of the dimerized
Heisenberg or $XY$ spin chains.\cite{Bulaev63,Holi01}.
In the ordered AFE phase it shows the spin-gap behavior. At low
temperatures the asymptotics of the integral (\ref{chiS})
is given by the following expression
\begin{equation}
\label{chiSas}
\chi_s = \frac{1}{J+\lambda m_z^2} \Big(
\frac{2 \Delta_{\textrm{SG}}}{\pi T}
\Big)^{\frac12}
\textrm{exp} \Big[ -\frac{\Delta_{\textrm{SG}}}{T} \Big]
\Big\{ 1+\frac38 \frac{T}{\Delta_{\textrm{SG}}}+
\mathcal O \Big( \frac{T^2}{\Delta_{\textrm{SG}}^2} \Big) \Big\}
\end{equation}
The characteristic energy scale parameter $\Delta_{\textrm{SG}}$
defined as
\begin{equation}
\label{DeltaSG}
\Delta_{\textrm{SG}} \equiv 2 \varepsilon m_x~,
\end{equation}
is natural to call the spin gap. Such definition coincides with
the one stated in terms of a magnon.\cite{Bulaev63}
The latter is equal to the minimal energy needed to flip one spin,
and according to Eqs.(\ref{E},\ref{sz}), to the one-particle
fermion gap.
The simple relation (\ref{DeltaSG}) between the spin
gap and the alternating part of the exchange $b=2 \varepsilon m_x$
[cf. Eq.(\ref{MolF}d)] holds because the $XY$ model is
a free-fermion problem. Accounting for fermionic interactions
in the $XYZ$ model breaks (\ref{DeltaSG}) already at the mean-field
level.\cite{Bulaev63} Bosonization treatment of the
interacting Jordan-Wigner fermions\cite{CF79} reveals the power-law
relationship $\Delta_{\textrm{SG}} \propto b^{2/3}$ times
some log correction.\cite{BE81}

It is interesting to compare the ratio of the zero-temperature
spin gap $\Delta_{\textrm{SG}}^{\circ}$ and the critical temperature
in different regimes of couplings.
At large $g \gg g_{\lambda}$ this ratio
[cf. Eqs.(\ref{mz0P},\ref{mx0P},\ref{Tcas})]
\begin{equation}
\label{RatIs}
\frac{\Delta_{\textrm{SG}}^{\circ}}{T_c} \approx
\frac{4 \varepsilon}{g}
\end{equation}
is small according to the assumption (\ref{parange}) on the range of the
model parameters we are working with. Upon decreasing $g$ the ratio
increases. It subtly involves interplay of the model couplings,
and it is not easy to get its analytic form. At the critical coupling
$g=g^*$ (when $\lambda =0$) it can be found from
Eqs.(\ref{TcCP},\ref{mx0asy}), and roughly
\begin{equation}
\label{RatCr}
\frac{\Delta_{\textrm{SG}}^{\circ}}{T_c}
\sim \Big(\frac{8}{\pi J} \Big)^\frac12
\varepsilon^2 \vert \ln \varepsilon \vert^\frac32
\end{equation}
More accurate evaluations of Eqs.(\ref{TcCP},\ref{mx0asy}), as well as
direct numerical calculations show that within the parameter range
(\ref{parange}) the ratio $\Delta_{\textrm{SG}}^{\circ}/T_c$ does not
exceed 1.
In the BCS region ($g < g_{\lambda}$) the ratio is maximal, and
moreover, it is the universal constant
[cf. Eqs.(\ref{Tcas},\ref{mx0asy})]
\begin{equation}
\label{BCSratio}
\frac{\Delta_{\textrm{SG}}^{\circ}}{T_c}=\frac{\pi}{e^{\gamma}}
\approx 1.76, \quad \textrm{BCS regime}~
\end{equation}
of any BCS-type theory.\cite{AGD} Also, as we can see from
Eq.(\ref{mxExpas}), the temperature dependence of the spin gap in this
regime is exactly as that of the superconductivity gap provided by
the BCS theory:\cite{AGD}
\begin{equation}
\label{BCSgap}
\Delta_{\textrm{SG}} \approx 3.06 T_c \tau^{\frac12}
 ~~\tau \ll 1,~\textrm{BCS regime}~.
\end{equation}
%
%%%%%%%%%%%%%%%%%%%%%%%%%%%%%%%%%%%%%%%%%%%%%%%%%%%%%%%%%%%%%%%%%%%%%%%%%%%%%%
\subsection{Charge (pseudospin) susceptibilities}\label{chiCha}
%%%%%%%%%%%%%%%%%%%%%%%%%%%%%%%%%%%%%%%%%%%%%%%%%%%%%%%%%%%%%%%%%%%%%%%%%%%%%%
%
We will be interested in the pseudospin (i.e., charge) susceptibilities
with respect to two external (electric) fields which have the same spatial
dependence as the Weiss fields $h^{i}_{mn}$. Namely, the field along
$z$ pseudospin ``direction" is constant $E^z_{mn}=-E_z$, while the field
$E^x_{mn}=(-1)^{m+n+1}E_x$ is staggered. We define the charge
susceptibilities as
\begin{equation}
\label{chiChaDef}
\chi^c_{ij} \equiv \frac{\partial m_i }{ \partial E_j}
\Big \vert_{E_j=0}~: ~(i,j)=x~\textrm{or}~z
\end{equation}
These quantities can be calculated from equations of
Section \ref{MF} with the Weiss fields shifted as $h_i \mapsto h_i+E_i$.
%
%%%%%%%%%%%%%%%%%%%%%%%%%%%%%%%%%%%%%%%%%%%%%%%%%%%%%%%%%%%%%%%%%%%%%%%%%%%%%%
\subsubsection{$T>T_c$}
%%%%%%%%%%%%%%%%%%%%%%%%%%%%%%%%%%%%%%%%%%%%%%%%%%%%%%%%%%%%%%%%%%%%%%%%%%%%%%
%
After a straightforward but somehow lengthy algebra we obtain for the
disordered paraelectric (PE) phase $T>T_c$:
\begin{subequations}
\label{chiCPE}
\begin{eqnarray}
\chi^c_{xz} &=& \chi^c_{zx}=0 \\
\chi^c_{zz} &=& \frac{\beta}{4}
\frac{1-4m_z^2}{1- \frac{\lambda \beta}{2 \pi}
(1-4m_z^2)(t_n+\lambda \beta m_z \dot{t}_{n,\tilde \beta}) }\\
\chi^c_{xx} &=&\frac{m_z}{1-m_z \big(
 g+ \frac{4 \varepsilon^2}{\pi(J+\lambda m_z^2)} \eta_n
 -\frac{2\lambda }{\pi} t_n \big) }
\end{eqnarray}
\end{subequations}
where
\begin{equation}
\label{tnbe}
\dot{t}_{n,\tilde \beta} \equiv
\frac{\partial t_n(\Delta,\tilde \beta)}{\partial \tilde \beta}~.
\end{equation}
Note that even in the limit $E_z \to 0$, $\chi^c_{zz}$ is in
fact the susceptibility under the applied field $\Omega$
[cf. Eq.(\ref{Ham})] along this direction. It does not show
critical behavior. On the contrary, from the comparison
of Eqs.(\ref{mzbe},\ref{chiCPE}c) we immediately see
that $\chi^c_{xx}$ diverges when $T \to T_c^{+}$.
At large $g \gg g_{\lambda}$ from Eqs.(\ref{Tcas},\ref{mzas})
we find
\begin{subequations}
\label{chiCPEasIs}
\begin{eqnarray}
\chi^c_{zz} &\approx& \frac{1}{4T} \\
\chi^c_{xx} &\approx&  \frac{4T_c}{g^2-4} \frac{1}{|\tau|}
\end{eqnarray}
\end{subequations}
so $\chi^c_{xx}$ diverges with the mean-field value of the critical index.
The same divergence can be obtained from the analytical
treatment in the BCS regime $g < g_{\lambda}$ for $T \agt T_c \ll 1$
where
\begin{subequations}
\label{chiCPEasBCS}
\begin{eqnarray}
\chi^c_{zz} &\approx& \frac{1}{T}
\textrm{exp}\big[ -\frac{\Delta_{\textrm{CG}}}{T}  \big] \\
\chi^c_{xx} &\approx&
\frac{\pi \tilde J }{4 \varepsilon^2 \ln \frac{T}{T_c} }
 \approx \frac{\pi \tilde J }{4 \varepsilon^2} \frac{1}{|\tau|}
\end{eqnarray}
\end{subequations}
At these low temperatures one probes the $\Omega$-generated
charge gap $\Delta_{\textrm{CG}}=1+\lambda /\pi \equiv g_{\lambda}/2 $
in $\chi^c_{zz}$, renormalized (comparatively to its bare value 1)
via the coupling $\lambda$ by the dimer operator average $t$
[cf. Eq.(\ref{Dmn})].
%
%%%%%%%%%%%%%%%%%%%%%%%%%%%%%%%%%%%%%%%%%%%%%%%%%%%%%%%%%%%%%%%%%%%%%%%%%%%%%%
\subsubsection{$T<T_c$}
%%%%%%%%%%%%%%%%%%%%%%%%%%%%%%%%%%%%%%%%%%%%%%%%%%%%%%%%%%%%%%%%%%%%%%%%%%%%%%
%
We will not give here the closed expressions for the four susceptibilities
below $T_c$, since there are too cumbersome. The important properties of
these quantities are the following: $\chi^c_{zz}$ is continuous across the
transition, and $\chi^c_{xx} \propto 1/\tau$ is divergent when approaching
$T_c$ from below. In the regimes of strong Ising coupling and BCS,
$\chi^c_{xx}$ can be calculated analytically, showing that
\begin{equation}
\label{chixxBe}
\chi^c_{xx}(T \to T_c^-) = \frac12 \chi^c_{xx}(T \to T_c^+)
\end{equation}
as one should have expected in the mean-field theory. The transverse
pseudospin susceptibilities are ``diamagnetic" (indicating that the system
tends to preserve the psedospin vector's length) and show weaker
divergence:
\begin{equation}
\label{offdiag}
 \chi^c_{xz}= \chi^c_{zx} \approx  \left\{
                \begin{array}{ll}
-\frac{1}{g} \Big( \frac{2}{g^2-4} \Big)^{\frac12}
\tau^{-\frac12},& g \gg g_{\lambda} \\[0.3cm]
-\frac{\pi^2}{2 \sqrt{14 \zeta(3)}}\frac{\tilde J T_c}{\varepsilon^3}
\tau^{-\frac12},
   &\textrm{BCS regime}
                \end{array}
       \right.
\end{equation}
%
%%%%%%%%%%%%%%%%%%%%%%%%%%%%%%%%%%%%%%%%%%%%%%%%%%%%%%%%%%%%%%%%%%%%%%%%%%%%%%
\subsection{Specific heat capacity}\label{heat}
%%%%%%%%%%%%%%%%%%%%%%%%%%%%%%%%%%%%%%%%%%%%%%%%%%%%%%%%%%%%%%%%%%%%%%%%%%%%%%
%
The easiest way to obtain the specific heat capacity in our approach, is
to calculate it directly from the mean-field energy
(\ref{Emf})
\begin{equation}
\label{cv}
c=\frac{\partial {\mathcal E}_{\textrm{MF}}}{\partial T}
\end{equation}
(here, in fact $c=c_v$). As usual, at the second-order phase transition
temperature the specific heat undergoes a finite jump. For two regimes
of couplings $c$ can be easily calculated analytically.
At high temperatures $T \gg 1$ above $T_c$
\begin{equation}
\label{cbigT}
c \approx \frac{1}{4T^2}\big( 1+\frac{J^2}{2} \big)
\end{equation}
As one can see from the above equation, spins and pseudospins ($m_z$)
contribute to the specific heat in the same fashion. At large
$g \gg g_\lambda$, $T_c$ is also large. For this regime we find
\begin{equation}
\label{delcIs}
c^- -c^+ \approx \frac{g^2-4}{4gT_c}, ~g \gg g_{\lambda}
\end{equation}
where $c^\pm \equiv c(T\to T_c^\pm)$. Low temperatures ($T \ll 1$) within
the disordered PE phase are attainable only at smaller couplings
$g \alt g_{\lambda}$. In the BCS regime ($T > T_c$) we get
\begin{equation}
\label{cBCS}
c \approx \frac{\pi T}{3 \tilde J}+\mathcal O (T^3), ~\textrm{BCS regime}
\end{equation}
Since pseudospins $m_z$ are gapped, at these low temperatures their
contribution to the specific heat is exponentially suppressed as
$\mathcal O(\exp[-\Delta_{\textrm{CG}}/T])$, thus only spins contribute.
The latter contribution is equivalent to the one
of a degenerate free electron gas. Eqs.(\ref{mabs})
for the AFE phase in the BCS regime gives
\begin{equation}
\label{dmz}
\frac{\partial m_z}{\partial T} =-\frac{\partial m_x^2}{\partial T}
+\mathcal O(e^{-1/T})
\end{equation}
Using it along with asymptotics
(\ref{tnas},\ref{etanas},\ref{Tcas},\ref{aas},\ref{bas},\ref{mxExpas}),
we obtain
\begin{equation}
\label{delcBCS}
c^- -c^+ \approx \frac{4 \pi}{7 \zeta(3)} \frac{T_c}{\tilde J},
\quad \textrm{BCS regime}
\end{equation}
Then the ratio
\begin{equation}
\label{cRat}
\frac{c^-}{c^+}=1 + \frac{12}{7 \zeta(3)} \approx 2.43
\end{equation}
is another universal constant known from the BCS theory.\cite{AGD}
Note that the origin of this universality can be traced from
Eq.(\ref{dmz}), which results in that only gapped fermions contribute
to the heat capacity in the vicinity of $T_c$ in the BCS regime.

Since the specific heat jump is positive at any regime of couplings,
it indicates the free energy (\ref{FE}) decrease in the ordered AFE
phase $f_{\textrm{AFE}} -f_{\textrm{PE}} \propto -m_x^2$.

Upon approaching zero temperature the specific heat capacity decreases
exponentially for any coupling $g$. The decay scale is defined
by $\Delta_{\textrm{SG}}$ which is the smallest gap in the model
(the zero temperature charge gap varies from
$\Delta_{\textrm{CG}}=g_{\lambda}/2$ in the BCS regime to
$\Delta_{\textrm{CG}}=g/2$ in the regime of large $g$),
so $c \propto \exp [ -\Delta_{\textrm{SG}}/T ]$ similar to the
spin susceptibility.
%
%%%%%%%%%%%%%%%%%%%%%%%%%%%%%%%%%%%%%%%%%%%%%%%%%%%%%%%%%%%%%%%%%%%%%%%%%%%%%%
\section{Summary and Discussion}\label{Concl}
%%%%%%%%%%%%%%%%%%%%%%%%%%%%%%%%%%%%%%%%%%%%%%%%%%%%%%%%%%%%%%%%%%%%%%%%%%%%%%
%
We study the spin-pseudospin model comprised from the Ising Model in Transverse
Field (IMTF) Hamiltonian for pseudospins coupled
to the spin-$\frac12$ $XY$-Hamiltonian, on a triangular lattice.
The effective Hamiltonian, similar to ours, appears from analyses of
the quarter-filled ladder compound $\rm NaV_2O_5$, which can be described
by a set of spin and pseudospin variables, the latter representing
the charge degrees of freedom.
Following the proposal of Mostovoy and Khomskii\cite{Most98,Most02}
for the scenario of the phase transition in $\rm NaV_2O_5$,
we include into the Hamiltonian a specific inter-ladder spin-pseudospin
coupling term  allowed by symmetries, linear over the pseudospin
operators.

In the framework of the molecular-field approximation we explore
the phase diagram of the model and find that it
possesses two phases: the disordered phase without charge order or
spin gap; and the low-temperature phase containing both the
anti-ferroelectric (zigzag) charge order and spin gap
(without long-range spin order). Such ordered phase
is experimentally observed below the transition temperature
$T_c=34\,{\textrm K}$ in $\rm NaV_2O_5$.
The phase transition in our model is of the second kind.
We calculate various physical quantities, as
the order parameter, spin and pseudospin (charge) susceptibilities,
specific heat, and find their (mean-field) temperature behavior near
$T_c$.

Our analysis reveals an important property of the phase diagram,
not envisaged in the original Mostovoy-Khomskii proposal: the
inter-ladder spin-pseudospin coupling ($\varepsilon$) not only
creates the spin dimerization (spin gap)
triggered by the charge ordering in the region of Ising's coupling
where the IMTF can order ($g>g_{\lambda}$), but it also generates
simultaneous appearance of the charge order and spin gap in the case
when the IMTF is always disordered ($g<g_{\lambda}$).
So, the coupling $\varepsilon$ destroys the IMTF quantum critical
point at $g=g_{\lambda}$ and results in a
continuous evolution of $T_c(g) \neq 0$ into the region
$g<g_{\lambda}$. In that region $T_c$ has the exponential BCS-like
dependence on model's couplings.

An interesting feature of model's phase diagram is that near the IMTF
critical coupling $g_{\lambda}$ it shows regimes with a re-entrance
(for couplings $\lambda \neq 0$, $\varepsilon=0$) or with
a double re-entrance (for couplings $\lambda \neq 0$,
$\varepsilon \neq 0$). How pronounced the re-entrant behavior is,
depends on the interplay of couplings, but the re-entrance is absent
when $\lambda =0$. Usually, re-entrance is due either to disorder
effects,\cite{Kir00} or due to the stabilization of the ordered phase
via the release of entropy by secondary degrees of
freedom.\cite{Pietig99}
The IMTF does not show any re-entrance at the mean-field
level,\cite{Blinc74,Chak96} but accounting for the first-order
fluctuation corrections revealed some re-entrance in the
2D IMTF.\cite{Stratt86} In that study it was assumed to be an
artifact of poor approximation. In our case, we have given
some qualitative arguments on how the re-entrant behavior
can be understood from the competition of model's energy
scales. However, we cannot exclude that the found re-entrance
is a mean-field artifact. A more detailed study on this
re-entrance will be presented elsewhere.\cite{CGlet03}

\begin{figure*}
\includegraphics[width=12cm]{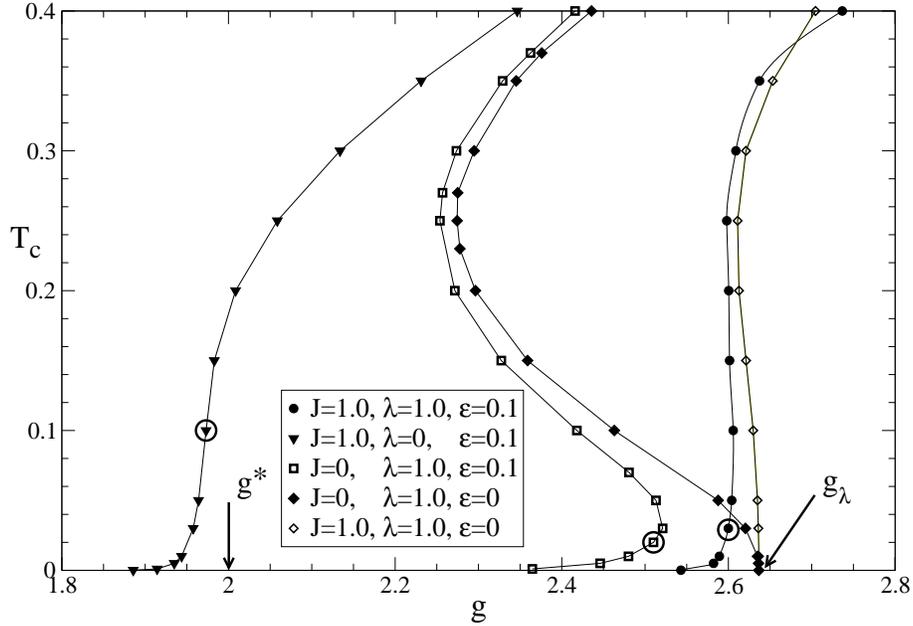}
\caption{
Critical temperature of the AFE phase transition as a function of the Ising
coupling $g$ at different
values of $\lambda, \varepsilon$. The critical couplings $g^*=2$ and
$g_{\lambda}=2.6366$. Large empty circles on the curves at
$\varepsilon =0.1$ indicate the right boundary for the BCS-region
described by the exponential dependence (\ref{Tcas}).
At large values of $g$ (not shown) all curves $T_c(g)$
approach the asymptotic line $T_c =g/4$.
}
\label{TcFig}
\end{figure*}

Now we proceed with the discussion on the applications of our results
to $\rm NaV_2O_5$.
For $\rm NaV_2O_5$ we take the following data:\cite{Smo98,Fagot00,Lem03}
$\Omega =700$ meV, $T_c=34$ K,
$\Delta_{\rm SG}=106$ K ($\Delta_{\textrm{SG}}^{\circ}/T_c \approx 3$),
$J_H+\frac14 J_{ST}=50$ meV. The microscopic calculations\cite{GrosEps}
give $\tilde \varepsilon= 15$ meV. Using these data as
model's dimensionless parameters [cf. notations (\ref{Coupdls})]
$\varepsilon=0.021$, $J=0.065$,
$\lambda=0.026$ we find that $T_c=0.004$ lies on the curve $T_c(g)$
in the BCS-regime region with $g=1.99$ ($J_I=1.4$ eV),
close to the critical coupling $g_{\lambda}=2.02$. The estimates for
$J_I$ in $\rm NaV_2O_5$ which is proportional to the in-ladder Coulomb
repulsion between neighboring rungs, vary,\cite{Smo98,Most98,Deb00}
but do not exceed $J_I=1.5$ eV.\cite{Ohta03} The whole curve $T_c(g)$
with the above parameters $J,\lambda,\varepsilon$ looks very much
similar to that shown in Fig. \ref{TcFig} for $J=1,\lambda=0,
\varepsilon=0.1$, without appreciable re-entrance. (In any case
$g=1.99<g_{\textrm{min}}$.) We would not insist that our mean-field
analysis gives the quantitatively correct evaluation of $T_c$ in
terms of the microscopic parameters for $\rm NaV_2O_5$, since the
uncertainties in their values left us with a freedom to make this
reasonable fit. However, we think we have grasped the qualitatively
correct physics. The question why $T_c$ is so low in $\rm NaV_2O_5$,
or what is the scale that gives such a low $T_c$, can be answered
as follows: $g$ for $\rm NaV_2O_5$ lies in the proximity of
the quantum critical point of the IMTF (most likely on the ``disordered
side, i.e., $g< g_{\lambda}$). $T_c$ is determined by the two
scales: $g-g_{\lambda}$ and the inter-ladder dimerization coupling
$\varepsilon$. The latter destroys the IMTF quantum critical point
and makes the ordering possible at $g< g_{\lambda}$.

A flaw of the present study is that we cannot account for the large
BCS ratio $\Delta_{\textrm{SG}}^{\circ}/T_c$ observed in
$\rm NaV_2O_5$.
The small value of $T_c$ indicates that we are either in the
BCS regime described by the analytical result (\ref{Tcas}), or very
close to it, so the ratio cannot exceed the universal BCS value
(\ref{BCSratio}). This gives the spin gap almost two times smaller
than the experimental value.  This clearly indicates that
for more realistic calculations the spin part of the Hamiltonian
should be modified, and instead of the $XY$ spin model the full
three component Heisenberg Hamiltonian should be considered.
As we have already explained it in Section \ref{chiSpin},
this will change results for the BCS ratio. At this
point we cannot say whether this will be enough in order
to explain the experimentally observed magnitude of the
 ratio.
Another possibility
is of course, the role of modes, neglected in our model,
like, e.g., phonons, which can affect the BCS ratio. However,
we will not speculate more on this point.

Experiments of various kinds\cite{Lem03} unequivocally
demonstrate that what occurs in $\rm NaV_2O_5$ at $T_c =34$ K
is a (thermal) continuous structural phase transition.
Also rather large regions of structural (charge-ordering)
two-dimensional fluctuations are seen\cite{Gaulin00} on the both
sides of $T_c$. The order parameter critical
index\cite{Ravy99,Gaulin00,Fagot00}
$\beta \approx 0.17-0.19$ is close to what one expects for a
two-dimensional phase transition with a one-component order parameter.
So we conclude that the universality class of this transition is
the 2D Ising. In order to obtain a more complete description
of the transition in $\rm NaV_2O_5$,
thermal fluctuations in the IMTF-sector
of the spin-pseudospin Hamiltonian must be taken into account.
According to what we said above, we expect the Ising coupling
for the $\rm NaV_2O_5$-Hamiltonian to lie near the quantum
critical point of the 2D IMTF. From the experimental results
we expect that physics of $\rm NaV_2O_5$ near $T_c$ is controlled
by the region of the 2D IMTF phase diagram,\cite{Friedman78}
where 2D Ising (thermal) fluctuations dominate. Although
at lower temperatures the change of the nature of these fluctuations
into 3D Ising (quantum),\cite{Friedman78} may also become
important. We believe that apart from the applications to
$\rm NaV_2O_5$, a study of the fluctuations in proximity
of this quantum critical point is a very interesting problem
of its own.

\begin{figure*}
\includegraphics[width=7cm]{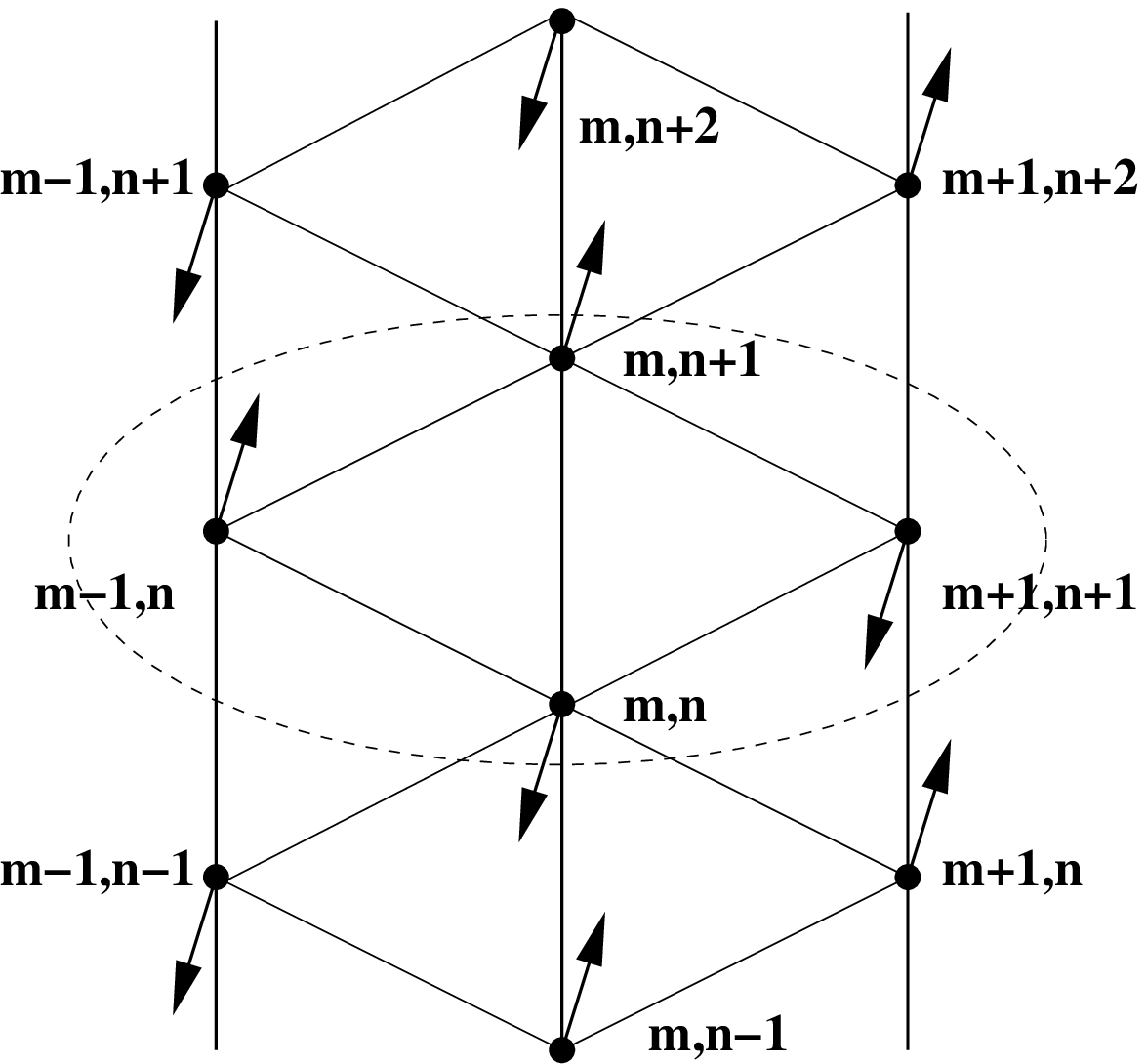}
\caption{
Two-dimensional triangular lattice of coupled ladders. A vertical line
denotes a single ladder where a dot denotes its rung. At each rung $(m,n)$
there is pseudospin $\bm{\mathcal T}_{mn}$ and spin $\mathbf S_{mn}$ (not
shown). In the region encirled by the dashed line we show four sites
involved in each term of the sum in the Hamiltonian (\ref{Hamdls}). Spins
and pseudospins residing on two sites of the same $m$-th ladder are coupled
via the exchange interaction terms, while two pseudospins from $(m-1)$-th
and $(m+1)$-th ladders are coupled via the dimerization interaction
constant $\varepsilon$. The pseudospin $\mathcal T_{mn}^x$
ordering pattern shown in the
figure corresponds to the ordered anti-ferroelectric (zigzag) phase.
}
\label{LatFig}
\end{figure*}

It is worth pointing out that
the pseudospin sector of our Hamiltonian is only \textit{quasi}
two-dimensional, in the sense that pseudospins on neighboring
ladders are coupled only via the dimerization coupling $\varepsilon$.
Accounting for the Ising coupling between ladders, as one can see
from the triangular lattice mapping shown in Fig.\ref{LatFig},
will cause a frustration. The neglect of this coupling for the
applications to $\rm NaV_2O_5$ is justified from microscopic
grounds, but from a general point of view it would be interesting
to study a Hamiltonian, similar to ours, with a truly two-dimensional
IMTF part.
%
%%%%%%%%%%%%%%%%%%%%%%%%%%%%%%%%%%%%%%%%%%%%%%%%%%%%%%%%%%%%%%%%%%%%%%%%%%%%%%
%%%%%%%%%%%%%%%%%%%%%%%%%%%%%%%%%%%%%%%%%%%%%%%%%%%%%%%%%%%%%%%%%%%%%%%%%%%%%%
%------------------------------------------------------------------------------
\begin{acknowledgments}
%\acknowledgements
We are grateful to E. Orignac and R. Valent\'\i\ for helpful discussions.
We are also thankful to F. Capraro and K. Pozgajcic for their
help with the software and numerical calculation.
This work is supported by the German Science Foundation.
\end{acknowledgments}
%
%%%%%%%%%%%%%%%%%%%%%%%%%%%%%%%%%%%%%%%%%%%%%%%%%%%%%%%%%%%%%%%%%%%%%%%%%%%%%%
% REFERENCES
%%%%%%%%%%%%%%%%%%%%%%%%%%%%%%%%%%%%%%%%%%%%%%%%%%%%%%%%%%%%%%%%%%%%%%%%%%%%%%

%xxxxxxxxxxxxxxxxxxxxxxxxxxxxxxxxxxxxxxxxxxxxxxxxxxxxxxxxxxxxxxxxxxxxxxxxxxxxxx
%xxxxxxxxxxxxxxxxxxxxxxxxxxxxxxxxxxxxxxxxxxxxxxxxxxxxxxxxxxxxxxxxxxxxxxxxxxxxxx

%xxxxxxxxxxxxxxxxxxxxxxxxxxxxxxxxxxxxxxxxxxxxxxxxxxxxxxxxxxxxxxxxxxxxxxxxxxxxxx
%xxxxxxxxxxxxxxxxxxxxxxxxxxxxxxxxxxxxxxxxxxxxxxxxxxxxxxxxxxxxxxxxxxxxxxxxxxxxxx
%xxxxxxxxxxxxxxxxxxxxxxxxxxxxxxxxxxxxxxxxxxxxxxxxxxxxxxxxxxxxxxxxxxxxxxxxxxxxxx

\begin{thebibliography}{}
\bibitem{PnNote} Contrary to $\rm NaV_2O_5$, the quasi-2D compound
$\rm Na_2Ti_2Pn_2O$ is less studied experimentally and less
understood,\cite{Pn2O} so in the following we will restrict our
discussion to the former.
\end{thebibliography}
\end{document}